\documentclass[]{raa}            
\usepackage{graphicx,times}
\usepackage{natbib}

\providecommand{\bm}[1]{\mbox{\boldmath$#1$\unboldmath}}

\begin{document}

\title{On the origin of X-shaped radio galaxies}

 \volnopage{ {\bf 2011} Vol.\ {\bf 11} No. {\bf XX}, 000--000}
   \setcounter{page}{1}

\author{Gopal-Krishna
  \inst{1}
\and Peter L.\ Biermann
  \inst{2,3,4,5,6}
\and L\'{a}szl\'{o} \'{A}. Gergely
  \inst{7,8}
\and Paul J.\ Wiita
  \inst{9}
}

\institute{National Centre for Radio Astrophysics, TIFR, Pune University Campus, Post Bag 3, Ganeshkhind, Pune 411007, India; {\it  krishna@ncra.tifr.res.in}\\
  \and
Max Planck Institute for Radioastronomy, Auf dem H{\"u}gel 69, 53121 Bonn, Germany; {\it plbiermann@mpifr-bonn.mpg.de}\\
  \and
Department of Physics \& Astronomy, University of Bonn, Germany\\
  \and
Department of Physics \& Astronomy, University of Alabama, Tuscaloosa, AL, USA\\
  \and
Department of Physics, University of Alabama at Huntsville, AL, USA\\
  \and
FZ Karlsruhe, and Physics Department, University of Karlsruhe, Germany\\
  \and
Department of Theoretical Physics, University of Szeged, Tisza Lajos k{\"o}r{\'u}t 84--86, 6720 Szeged, Hungary; {\it gergely@physx.u-szeged.hu}\\
  \and
Department of Experimental Physics, University of Szeged, D\'om t\'er 9, 6720 Szeged, Hungary\\
  \and
Department of Physics, The College of New Jersey, P.O.\ Box 7718, Ewing, NJ 08628-0718, USA;
{\it wiitap@tcnj.edu}
}
\vs \no
   {\small Received [year] [month] [day]; accepted [year] [month] [day] }

\abstract
{After a brief, critical review of the leading explanations proposed for the 
small but important subset of radio galaxies showing an X-shaped morphology 
(XRGs) we propose a generalized model, based on the jet-shell interaction and 
spin-flip hypotheses. The most popular scenarios for this intriguing phenomenon 
invoke either hydrodynamical backflows and over-pressured cocoons or rapid jet 
reorientations, presumably from the spin-flips of central engines following the 
mergers of pairs of galaxies, each of which contains a supermassive black hole 
(SMBH).  We confront these models with a number of key observations and thus
argue that none of the models is capable of explaining the entire range 
of salient observational properties of XRGs, although some of the 
arguments raised in the literature against the spin-flip scenario are 
probably not tenable. We then propose here a new scenario which also involves 
galactic mergers but would allow the spin of the central engine to maintain its 
direction. Motivated by the detailed multi-band observations of the nearest 
radio galaxy, Centaurus A, this new model emphasizes the role of interactions 
between the jets and the shells of stars and gas that form and rotate around the merged 
galaxy and can cause temporary deflections of the jets, occasionally giving 
rise to an X-shaped radio structure.   Although each of the models is likely 
to be relevant to a subset of XRGs, the bulk of the evidence indicates that most 
of them are best explained by the jet-shell interaction or spin-flip hypotheses.
}
 \authorrunning{Gopal-Krishna, Biermann, Gergely \& Wiita }            
   \titlerunning{Origin of X-shaped Radio Galaxies }  
   \maketitle

\hyphenation{mono-chro-matic  sour-ces  Wein-berg
chang-es Strah-lung dis-tri-bu-tion com-po-si-tion elec-tro-mag-ne-tic
ex-tra-galactic ap-prox-i-ma-tion nu-cle-o-syn-the-sis re-spec-tive-ly
su-per-nova su-per-novae su-per-nova-shocks con-vec-tive down-wards
es-ti-ma-ted frag-ments grav-i-ta-tion-al-ly el-e-ments me-di-um
ob-ser-va-tions tur-bul-ence sec-ond-ary in-ter-action
in-ter-stellar spall-ation ar-gu-ment de-pen-dence sig-nif-i-cant-ly
in-flu-enc-ed par-ti-cle sim-plic-i-ty nu-cle-ar smash-es iso-topes
in-ject-ed in-di-vid-u-al nor-mal-iza-tion lon-ger con-stant
sta-tion-ary sta-tion-ar-i-ty spec-trum pro-por-tion-al cos-mic
re-turn ob-ser-va-tion-al es-ti-mate switch-over grav-i-ta-tion-al
super-galactic com-po-nent com-po-nents prob-a-bly cos-mo-log-ical-ly
Ex-tra-po-la-ting Kron-berg Berk-huij-sen stra-ti-fied puzz-ling nu-cle-us dis-cov-ered} 

\section{Introduction}

Morphologies of powerful extragalactic radio sources on kiloparsec and
larger scales can typically be related to a single pair of anti-parallel
jets of relativistic plasma, ejected from a \textquotedblleft central
engine" located at the active nucleus of a massive early-type galaxy (e.g.,
Begelman, Blandford \& Rees 1984) . In some rare cases, an additional inner
pair (or pairs) of radio lobes is observed, roughly aligned with the outer,
presumably older, lobe pair. These, so called, \textquotedblleft
double-double" radio galaxies are believed to be examples of restarted
nuclear activity (Lara et al.\ 1999; Schoenmakers et al.\ 2000; Saripalli,
Subrahmanyan \& Udaya Shankar 2002; Konar et al.\ 2006), 
in which the new jet pair broadly retains its alignment with
the older jets.   In one scenario, the interruption in the jet activity is attributed
to the disruption of the accretion disk due to the inspiral of a second
supermassive black hole through the disk surrounding the larger one (Liu et al. 2003).
In the last few years an even more striking morphological 
class of radio galaxies has drawn increasing attention; they exhibit two 
pairs of radio lobes associated with a single parent galaxy, such that 
their axes are grossly misaligned. Although the existence of such 
\textquotedblleft winged" or \textquotedblleft X-shaped" radio galaxies 
(XRGs) has been known for the past few decades (e.g., H\"{o}gbom \& Carlsson 
1974; Ekers et al.\ 1978; Leahy \& Williams 1984; Leahy \& Parma 1992), 
the recent upsurge of interest in XRGs stems from the distinct possibility 
that their precursors were potential sites of the intense gravitational 
radiation which is expected in a merger of the two supermassive black holes 
(SMBHs) associated with the merged galaxy pair.  Another simultaneous 
outcome expected from such a merger is a sudden change in the spin axis of 
the SMBH, which can find manifestation in the  emergence of a new jet pair 
at an axis grossly misaligned from the axis of the jets prior to the merger, 
producing the so called, \textquotedblleft spin-flip" model for XRGs (e.g., Rottmann 
2001; Zier \& Biermann 2001, 2002; Merritt \& Ekers 2002; Biermann et al.\ 
2005; Gergely \& Biermann 2009, hereafter GB09). An earlier explanation, specifically devised for the rare dumbbell 
galaxies (elliptical galaxies with two nuclei, such as the XRG NGC 326), 
posited a more gradual reorientation of the large-scale jets during a close 
passage between two elliptical galaxies (Wirth, Smarr \& Gallagher 1982). 

The explanation invoking a SMBH merger accords well with the extensively
discussed model of powerful double radio sources that argues that jet
formation is often triggered by a galaxy merger, particularly the merger of
a gas-rich galaxy with a large elliptical (e.g., Begelman et al.\ 1984;
Heckman et al.\ 1986; Blandford 1992; Wilson \& Colbert 1995; Wang \&
Biermann 1998).   The possibility that inversion symmetric radio galaxy morphologies
were due to a SMBH binary in a single galactic nucleus was first suggested by Begelman et al.1980).  Significant additional support for this scenario comes from
the growing evidence in favor of the presence of binary black hole systems
within the nuclei of active galaxies. One line of evidence comes from recent
discoveries of double-peaked broad low-ionization lines separated by $> 2000$
km s$^{-1}$ (Komossa, Zhou \& Lu 2008; Boroson \& Lauer 2009; Shields, et
al.\ 2009); however, it must be noted that these systems could just
be superpositions of two active galactic nuclei (AGN) within the same
cluster of galaxies (e.g., Dotti \& Ruszkowski 2009).  X-ray imaging and spectroscopy
indicate that both the optical nuclei (separated
by a projected distance of $\sim$1.3 kpc) of the ultraluminous infrared galaxy NGC 6240 
are active and thus house SMBHs (Komossa et al.\ 2003).   It has been noted that
the radio emission from NGC 1275 is consistent with accelerating jet precession and
this could imply the presence of a binary SMBH system with a rapidly changing orbital separation
(Liu \& Chen 2007).  More interestingly,
double-peaked low-ionization emission lines in the nucleus of a galaxy
hosting an XRG J1130+0058 have been reported  (Zhang, Dultzin-Hacyan 
\& Wang 2007).    Very recently, the detection of a galaxy containing three active nuclei, SDSS J1027$+$1749, has been reported (Liu et al.\ 2011);
here the secondary emission line nuclei are offset by 2.4 and 3.0 kpc in projected separation and by 450 and 111 km s$^{-1}$ in velocity.
Further evidence for binary SMBHs comes from models of the
periodic intensity variations of some blazars, notably OJ 287 
(Sillanp{\"a}{\"a} et al.\ 1988; Valtonen et al.\ 2008). The detection in the
RG 0402$+$379 of two compact, VLBI sources of variable radio emission, 
separated by only 7.3 pc, also presents a strong case for a SMBH merger in 
progress (Rodriguez et al.\ 2006).   An earlier stage leading to an eventual merger
is very possibly exemplified by the highly disturbed radio morphology in the compact steep radio
spectrum quasar, FIRST J164311.3$+$315618, which is in a binary system with
a projected separation of 15 kpc (Kunert-Bajraszewska \& Janiuk 2011).

The slow accumulation of observational evidence notwithstanding, the
occurrence of SMBH mergers has been long firmly believed on the basis of simple
theoretical arguments. For instance, the Press-Schechter model (Press \& 
Schechter 1974) is
premised on an ongoing merger process for galaxy formation and this yields
a good approximation to the galaxy luminosity function. When this model (or 
more sophisticated cosmological simulations) is coupled with the more recent
paradigm that a very large fraction of galaxies undergo central activity 
emitting a compact radio signature (e.g. Perez-Fournon \& Biermann 1984; 
Nagar et al.\ 2001; Nagar, Falcke \& Wilson 2005; Capetti et al. 2010) and 
that most galaxies harbor a central BH (e.g., Kormendy \& Richstone 1995; 
Faber et al.\ 1997; Magorrian et al.\ 1998), the expectation is that SMBH 
mergers are not uncommon.

XRGs are now known to constitute a significant fraction (up to $10\%)$ of
the more powerful, edge-brightened, FR II (Fanaroff \& Riley 1974) radio
galaxies in the 3CRR catalog (see, Leahy \& Williams 1984, hereafter LW84;
Leahy \& Parma 1992). Many new XRGs, or at least good XRG candidates, have 
recently been found and characterized (e.g., Cheung 2007, hereafter C07; 
Cheung et al.\ 2009; Saripalli \& Subrahmanyan 2009, hereafter SS09), mostly 
using the FIRST radio survey (Becker, White \& Helfand 1995). Usually XRGs 
are found to have radio powers near to the radio luminosity of the Fanaroff 
\& Riley (1974) division between FR I and FR II sources at 
$P_{178 \mathrm{MHz}} \approx 2 \times 10^{25}$ W Hz$^{-1}$ sr$^{-1}$
(Landt, Cheung \& Healey 2010 and references therein). 
Therefore it is striking that not a single case is found where both lobe 
pairs are of FR II type; usually only one (but sometimes neither) of the 
two lobe pairs shows radio hot spots near the lobes' extremities, while the 
other lobe pair lacks compact features altogether and normally is distinctly less
collimated (e.g., LW84; SS09). That pair is therefore termed as 
\textquotedblleft secondary lobes", or \textquotedblleft wings", in contrast 
to the other lobe pair, called \textquotedblleft primary lobes".

Several authors have argued that the wings arise due to diversion of the 
backflow of synchrotron plasma in the two primary lobes. The diversion
could be preferrentially into the cavities left over by the lobes of a
previously active phase (LW84), or due to buoyancy pressure driven by a 
steep pressure gradient in the ambient medium (e.g.,  Worrall, Birkinshaw 
\& Cameron 1995; Kraft et al.\ 2005; Miller \& Brandt 2009; SS09; Hodges-Kluck 
et al. 2010b). Another 
variant to this buoyant backflow model is the \textquotedblleft over-pressured 
cocoon" model where a rapid build-up of the pressure within the radio cocoon 
leads to a collimated supersonic outflows of the lobe's synchrotron plasma 
as it squirts out along the directions of the fastest declining ambient 
pressure (Capetti et al.\ 2002, hereafter C02).

In a less discussed variant of this class of models (Gopal-Krishna, Biermann
\& Wiita 2003, hereafter GBW03), the backflow gets diverted as it impinges,
from opposite sides, on to an inclined `superdisk' of thermal plasma around
the host elliptical. The existence of such superdisk structures, with mean
thickness of $\approx$30 kpc, around a number of FR II RGs has been inferred
from the sharp strip-like emission gaps seen between their radio lobe pairs
(Gopal-Krishna \& Wiita 2000a, 2009). There are a number of physically reasonable models 
for the formation of the superdisks (e.g., Gopal-Krishna, Wiita \& Joshi 2007; 
Gopal-Krishna \& Wiita 2009). One model worth highlighting in the present 
context considers the physics of the final stage of the spin-flip of merging 
BHs, which is marked by a rapid precession of the BH axis due to which the 
jets swing about rapidly (GB09). 
These swinging jets can produce a narrowing cone-like wind through fast and 
massive entrainment of the surrounding gas, thus potentially giving rise to a 
superdisk (GB09). If true, this picture could be useful in identifying the
best candidates for an imminent merger of SMBHs.

In Section 2 we contrast the different explanations proposed for the XRG
phenomenon with the current observational results.  In Section 3 we propose a 
new additional mechanism enriching the phenomenology of galaxy-galaxy 
interactions when one or both of the galaxies harbor a central super-massive 
black hole: {\it the jet-shell interaction model} for XRGs. Section 4  
summarizes our main conclusions. 

\section{Models for XRGs}

In this section we briefly recapitulate the three main explanations,
including some of their variants, that have been proposed for the XRG 
phenomenon. 

First, however, we summarize some key results concerning the orientations
of radio emission axes and optical shells with respect to the optical 
shape of their host galaxies.  For many years, only rather small samples of RGs 
($\sim$ 100) could be used and, while there were claims of a correlation 
between radio emission and the minor axes of the host elliptical galaxies
(e.g., Palimaka et al.\ 1979), no significant correlation was found in
some other studies (e.g., Sansom et al.\ 1987). 

An important result has 
come from the recent work of Battye \& Browne (2009) who employed a much 
larger sample of $>$14,000 galaxies obtained from matching the FIRST radio 
survey with the SDSS optical catalog. They have demonstrated that late type 
galaxies bear an expected strong correlation between the optical major 
axis and the radio emission axis, presumably arising from the association 
of the radio emission with star formation in their disks.  On the other 
hand, for early type galaxies they do find a significant correlation 
between the radio axis and the optical minor axis; however, this correlation 
is dominated by the galaxies with lower ratios of radio to optical powers. 
For stronger radio-loud ellipticals, with which we are concerned here, 
no significant correlation of radio axis with optical axis was found (Battye 
\& Browne 2009). 

At the same time, as shown in Capetti et al.\ (2002), XRGs are only hosted by substantially
elongated ellipticals (ellipticity, $\eta~ \lower 2pt \hbox {$\buildrel >
\over {\scriptstyle \sim }$}~ 0.2$). 
In addition, C02 found that the principal radio axis defined by the two hot
spots of the primary lobe pair shows a distinct tendency to be close to the
optical major axis of the parent elliptical. Perhaps an even tighter
alignment exists between the host galaxy's minor axis and the secondary radio
axis as defined by the pair of the wings (C02).  Both of these trends have
been confirmed recently by SS09 and H-K10a using larger datasets. The
latter authors inferred the minor axis from the distribution of hot, X-ray
emitting gas detected with Chandra around the ellipticals in their sample. 
Hence for the XRG subset of stronger RGs, all studies agree 
that the primary radio lobe pair is preferentially oriented toward the 
major optical axis of the host elliptical (C02; SS09; Hodges-Kluck et al.\ 
2010a, hereafter H-K10a), though the samples on which this conclusion is 
based are still comparatively small. 

In certain cases a Z-symmetric morphology 
of the secondary lobes was observed (i.e, a lateral offset between the ridge lines of 
the two secondary lobes; see, Gopal-Krishna, Biermann \& Wiita 2003).   We later
show how such sources can provide a powerful discriminant between different XRG models.

Another more general important result, but one we believe
could be particularly relevant for  
XRGs, concerns optical shells around elliptical galaxies.  These 
optical shells, which probably arise during the course of a merger of an 
elliptical with a disky galaxy (e.g., Quinn 1984), are preferentially 
located along the major axes of post-merger ellipticals (e.g., Malin \& Carter 
1983; Sect.\ 3).

Important evidence in 
favor of mergers playing an important role in XRGs comes from a 
recent study of 29 XRGs which were recently compared to a control sample 
of 36 RGs with ``normal'' morphologies but similar redshifts and optical 
luminosities (Mezcua et al.\ 2010). These authors find that the members 
of the XRG sample have, on average, significantly more massive SMBH, as
would be expected if a merger engendered the change in jet direction 
in XRGs.  In addition, more direct evidence for merger has recently come
from the recent detection of shells in two XRGs, 3C 403 (Almeida et al.\ 2011) 
and 4C$+$00.58 (Hodges-Kluck et al.\ 2010b). 
 
\subsection{``Twin-AGN" Model for XRGs}

This model is inspired by the existence of elliptical galaxies with double
nuclei, e.g., NGC 326 (Battistini et al.\ 1980; Worrall et al.\ 1995).
In this model the pair of twin radio lobes is considered to be two
independent radio doubles associated with a close pair of active SMBHs
inside a merging pair of massive ellipticals (Lal \& Rao 2007).
This model appears particularly appealing in case the dynamical 
friction  is insufficient, thereby stalling the approach of the two SMBHs. 
However, such stalling is normally not expected (e.g., Zier 2007; Sesana, 
Hardt \& Madau 2007; Hayasaki 2009; GB09, and 
references therein), as also inferred from extensive VLBI imaging 
observations (Burke-Spolaor 2011).   In addition, this situation has a very
low probability, in that it requires both central engines to be simultaneously
launching jets.

While this ``twin AGN" picture has the advantage of 
conceptual simplicity, it fails to explain why none of the XRGs has both 
its lobe pairs of the FR II type. Nor does it explain the preference of
the primary lobe to be oriented towards the major axis of the 
optical host galaxy (C02; SS09; H-K10a). Furthermore, it is particularly 
hard in this picture to understand the often observed Z-symmetric morphology 
of the secondary lobes (Gopal-Krishna, Biermann \& Wiita 2003). On
all these grounds, it is evident that while this model might account for
a tiny fraction of XRGs, it is untenable for the bulk of the XRG population.

\subsection{Back-flow Diversion Models}

In the backflow diversion model for the formation of secondary lobe pair,
(e.g., LW84, Worrall et al.\ 1995; Kraft et al.  2005), the (denser) 
ISM of the host elliptical galaxy plays a key role by exerting buoyancy 
pressure on the backflowing synchrotron plasma within the two primary lobes. 
As a result, the synchrotron plasma is diverted away from the lobe axis, 
along the direction of the fastest declining ISM pressure. In this way, the 
strong tendency for the wings to align with the optical minor axis of the 
host elliptical can be readily understood. Although appealing for its 
simplicity, as noted by several authors (e.g., Dennett-Thorpe et al.\ 2002), 
this hydrodynamic backflow model is severely challenged by the key observation 
that in several XRGs the wings are found to be distinctly \textit{more extended} 
than the primary lobes, even though the latter are supposed to be advancing 
supersonically with respect to the external medium (whose ram pressure 
causes a bright hot spot marking the leading edge of the lobe), in contrast
to the subsonically growing wings. Examples of such XRGs can be found, e.g., 
in Leahy \& Parma (1992); SS09 and Sect.\ 4.

To circumvent this problem, C02 proposed that the wings can form due to
lateral expansion of the over-pressured cocoon of the double radio source,
along the steepest gradient in the ISM pressure (i.e., roughly along the
host galaxy's minor axis, assuming that the galaxy has high ellipticity,
making its ISM asymmetric). The lateral expansion of the synchrotron 
plasma could well occur in the form of a pair of loosely collimated supersonic 
flows of the synchrotron plasma out of the over-pressured cocoon, via a 
``de Laval nozzle'' formation (e.g., Blandford \& Rees 1974\footnote{We note that this physical concept of a spherical explosion in a stratified 
atmosphere leading to a jet-like formation of flow goes back to Kompaneets 
[1960] and was discussed in detail by Zeldovich \& Raizer [1966].}.) This mechanism 
for the formation of secondary lobes can possibly allow them to grow even 
longer than the primary lobes, as argued by C02 based on
two-dimensional hydrodynamical simulations. As these simulated ``outflows'' may be enhanced 
by the imposed cylindrical symmetry, the result 
needed to be verified by the use of three-dimensional (3D) simulations. 
Very recently, such 3D simulations have been performed and they showed
that under rather special circumstances such outflows with secondary lobes
with lengths comparable to those of the primary lobes could indeed occur
(Hodges-Kluck \& Reynolds 2011).
Furthermore, given the appropriate viewing angle, such secondary lobes could
appear to have a larger projected length than the primary ones (Hodges-Kluck \& Reynolds 2011;
Wiita, Sobczak \& Starr 2011).

Secondly, in this scenario, a 
large misalignment between the primary and secondary lobe pairs (a prerequisite 
for an XRG classification) would be realizable only in those cases where the 
jets happen to be ejected roughly along the optical major axis of the parent 
elliptical galaxy. {\it This selection bias could possibly force the backflow 
diversion model to be consistent with the observed tendency for the primary lobes 
in XRGs to be aligned with the optical major axis of the host galaxy}, for if 
the jets emerged roughly along the host's minor axis to begin with 
then any cocoon outflows would tend to coincide with the original jet direction 
itself and the source would not be classified as an XRG at all.

The main morphological features
of XRGs can be reproduced satisfactorily in this model. Nonetheless, it is
evident that such a scenario can only work provided the jets are not very
powerful, and therefore unable to exit quickly from the denser central region 
of the ISM, and possibly also from the intra-cluster medium (according to 
H-K10a),
before a substantial cocoon has developed. Indeed, numerical simulations of
more powerful jets crossing interstellar/intergalactic medium boundaries
show only modest sideways flows from the backflowing plasma and thus only
\textquotedblleft stubby\textquotedblright\ wings (e.g., Hooda \& Wiita
1996, 1998). Still, the over-pressured cocoon model does explain why radio
wings do not form in general and, more specifically, why they do not usually
form in the weaker, edge-dimmed, FR I sources: such RGs simply do not
generate the backflow needed to form over-pressured cocoons. Note that
recently SS09 have highlighted cases of a few XRGs where even the better
collimated (primary)
lobe pair appears to have the FR I morphology. They suggest that probably
these lobes too had an FR II past when their hotspots generated strong
backflows. Such a morphological transformation at later evolutionary stages
has been long predicted from dynamical considerations (e.g., Gopal-Krishna
\& Wiita 1988, 2000b; Gopal-Krishna 1991; Falle 1991; Bicknell 1994, 1995;
Kaiser \& Alexander 1997; Nakamura et al.\ 2008).

The concept of an over-pressured cocoon finds some support from a recent
study of the optical spectra of 53 XRGs, by Landt, Cheung \& Healey (2010)
who find that unusually hot plasma (T $>$ 15,000 K) commonly exists in the 
nuclear regions of XRGs. Still, it remains 
unclear if the over-pressure extends all the way to their cocoons. 
The difficulty is further underscored by the case of the XRG NGC 326 in 
which the wings are not only a few times longer than the primary lobes 
but the ``foot" of the eastern wing is located about 50 kpc from the nucleus, 
i.e., well outside the ISM of the galaxy (e.g., Murgia et al.\ 2001; Worrall 
et al.,\ 1995). This situation makes it hard to identify a mechanism to ensure
confinement of the hypothesized over-pressured cocoon.

Finally, we note that in support of the backflow origin of XRG wings (vis-{%
\`a}-vis the spin-flip scenario described in Section 2.3), SS09 have 
highlighted a few cases where an inner pair of radio lobes is seen fairly 
well aligned with the primary lobe pair, indicating that nuclear activity 
in those sources has restarted practically along the same axis. In 
Section 2.3 we shall propose alternative interpretations for this result.

\subsection{Rapid Jet Reorientation Models}

In this radically different scheme, the wings are envisioned to be relic
emission from the radio lobes created during an earlier phase of nuclear
activity when the jets were oriented in that direction. The jet orientation 
then underwent a rapid change, possibly an abrupt ``flip", resulting in the 
currently fed primary lobe pair. Models proposed for the jet reorientation 
invoke either precession or other realigning mechanisms (Ekers et al.\ 1978; 
Rees 1978; Klein et al.\ 1995; Dennett-Thorpe et al.\ 2002; Falceta-Gon\c{c}alves et al.\ 2010;
Hodges-Kluck et al. 2010b).     A particularly detailed model  
argues that the central SMBH is realigned with the binary orbital plane during
a process that involves a strong interaction between the binary and the accretion
disk that twists and warps the disk which quickly realigns the spin axis of the SMBH 
ejecting the jets (Liu 2004). 
Alternatively, the jet realignment can be a more violent process involving the 
coalescence of the jetted SMBH with another SMBH in the throes of a galaxy 
merger (Rottmann 2001; Zier \& Biermann 2001; Merritt \& Ekers 2002; 
Dennett-Thorpe et al. 2002; GBW03).
This latter mechanism is particularly interesting since, it also results in a copious 
emission of gravitational waves, which for the lower mass range 
of SMBHs, lies in the Laser Interferometer Space Antenna (LISA) frequency band. 

In an earlier paper (GBW03) we argued how the spin-flip scenario can also
readily account for the Z-symmetry about the nucleus, displayed by many of 
the wing-pairs. Pointing out the Z-distortion, GBW03 showed how it can arise 
prior to the spin-flip, as the jets propagated through the ISM of the massive
elliptical host. The outer portion of the ISM already had been set in slow 
rotation by the captured galaxy (roughly along its original orbital plane) as 
it spiraled into the radio-loud elliptical, eventually culminating in the 
coalescence of their black holes (see, also, Noel-Storr et al.\ 2003; Heinz 
et al.\ 2008). Thus, at large distances from the galactic nucleus the two
jets, having been slowed down, would be
diverted by the rotating ISM in opposite directions. The deflected flows 
can exist for extended periods and can arise at substantial distances from 
the source galaxy. Note that a recent study of a Z-shaped FR I RG, NGC 3801, 
provides strong evidence for the presence of a recent merger and a large fast 
rotating gas disk interacting with the jets (Hota et al.\ 2009). 

While the large misalignment between the two lobe pairs is statistically
expected in the spin-flip scenario, it might seem difficult to understand
the observed strong tendency for the primary lobes to align with the optical 
major axis of the host elliptical (C02; SS09; H-K10a). However, if the 
wings are forced into alignment with the optical minor axis due to the
buoyancy pressure of the ISM of the parent galaxy acting on the backflowing 
synchrotron plasma, the trend for the primary lobes to be 
broadly aligned with the optical main axis (Capetti et al. 2002; SS09) 
could follow simply from the requirement of a large spin-flip angle, or 
else the source would often end up being classified as a double-double RG, 
as e.g., in the case of J1453+3309 (Konar et al.\ 2006).

Evidence consistent with significant spin-flips comes from RGs that appear
to have restarted and have inner and outer structures that are significantly
misaligned. For example, 1448$+$63 shows  a nearly $90^{\circ}$ misalignment
between its pc and kpc structures (Giovannini et al.\ 2005), although such
differences can be exaggerated if the jet is relativistic and pointing close
to the line-of-sight. Less severe misalignments, but on much larger scales,
and hence not likely to be affected by such projection effects, are present 
in other sources. Examples include the giant radio galaxy J0116$-$ 473 
(Saripalli, Subrahmanyan \& Udaya Shankar 2002), J1453$+$3309 (Konar et al.\
2006) and 3C 293 (Evans et al.\ 1999).

In several XRGs an inner 
compact pair of radio lobes is seen and is roughly aligned with the primary lobe axis (see, SS09); 
this double-double morphology of the primary lobes, 
 can be easily reconciled with the spin-flip model and the usual expectation that double-double
RGs involve restarted jets (e.g. Lara et al.\ 1999).
Note that in a few of
the double-double XRGs, highlighted in SS09, the outer lobes lack hot
spots and these have been interpreted as the lobes which had an FR II
morphology previously (SS09). The possibility of large-scale jets/lobes
transforming from FR II to FR I at later stage in their lives has also been noted in
earlier studies (Gopal-Krishna \& Wiita 1988; Gopal-Krishna 1991; Falle 
1991; Bicknell 1994; Kaiser \& Alexander 1997; Nakamura et al. 2008;
Kawakatu et al. 2009).

A modified version of this scheme was proposed by Dennett-Thorpe et
al.\ (2002), based on their analysis of radio spectral variation across two
XRGs, 3C 223.1 and 3C 403. 
Although a careful analysis of photometric and spectroscopic evidence 
led Dennett-Thorpe et al.\ (2002) to conclude there is no evidence in the 
galaxy or its environs for a recent merger activity in either 3C 223.1 or 3C 403 
(see, also, Landt et al.\ 2010, for a similar assertion about a larger 
sample of XRGs), they point out that the backflow models
are not consistent with the relative sizes and spectra of the wings in 
these sources. Hence Dennett-Thorpe et al. suggest that the required 
rapid change in the jet axis within several Myr is either caused by a 
delayed SMBH merger following ingestion of a small galaxy by the radio
galaxy, which left no obvious signature in the latter's stellar population, 
or, alternatively, the axis flip may have occurred due to accretion disk 
instabilities (e.g., Natarajan \& Pringle 1998). 

In summary, the basic difference between the spin-flip and the backflow
diversion models is that  in the latter case both the primary and
secondary lobe pairs form and evolve quasi-simultaneously. However, in the spin-flip 
scenario only the secondary lobe pair (presently appearing as the wings)
existed prior to the SMBH spin-flip episode and the currently active primary 
lobe pair was created after the spin-flip.   

While discussing the three main approaches to model the XRG 
phenomenon, 
we have tried to underscore the difficulties each of those mechanisms 
encounters when confronted with the available range of observations. In 
Sect.\ 3 we shall consider a new scenario which appears to surmount the 
problems inherent to the existing models. Later, in Sect.\ 4 we shall 
contrast all four XRG models in a condensed tabular presentation. 

\section{A Jet--Shell Interaction Model for XRGs}

While this proposed new mechanism for XRG formation assumes that a galaxy 
merger has occurred, it does not necessarily require a spin-flip of the central
engine, in that the entire observed radio structure may have formed only 
after the merger is essentially complete. Originally mooted in GBW03, this 
mechanism is
premised on an interaction between the radio galaxy jets and stellar shells 
(e.g.,  Gopal-Krishna \& Chitre 1983) of the kind that are detected around
$\sim 10\%$ of nearby early-type galaxies located in low density environments
(Malin \& Carter 1983; Pierfederici \& Rampazzo 2004; Sikkema et al.\ 2007).
The shells are now also known to be fairly rich in gas (see below). 
Striking examples of jet-shell interactions on kiloparsec scale
have been noticed in the nearest radio galaxy, Centaurus A,  in particular, the
abrupt flaring of the Northern inner jet right at the point of its impact on a shell (Gopal-Krishna 
\& Saripalli 1984, hereafter GS84; Gopal-Krishna \& Wiita 2010, hereafter 
GW10).   In these papers it was also suggested that the S-shaped symmetry 
of the pattern of radio peaks about the galaxy is indicative a general clockwise
rotation of the shell complex.  A similar scenario, involving rotating shells
is proposed here for XRGs and can explain
the observed oppositely directed geometry of the wings.  
We have argued that such interactions in Cen A may be responsible 
for the production of many of the detected ultra-high energy cosmic rays 
(Gopal-Krishna et al.\ 2010).    In several nearby ellipticals radial sequences 
of such shells straddling the galaxy have been detected (e.g., Malin \& 
Carter 1980, 1983; Malin, Quinn \& Graham 1983; Prieur 1988; Sikkema et al.\ 
2007).  An elegant interpretation for these shells in terms of a phase-wrapping
formalism has been put forward, which invokes a merger of a disk galaxy with
a massive elliptical (Quinn 1984; Dupraz \& Combes 1986; Hernquist \& Quinn 
1988).

Clear signatures of such a merger indeed exist in the case of Cen A (see
Israel 1998, for a review). Since substantial quantities of HI and molecular
gas have since been detected in the Cen A shells (Schiminovich et al.\ 1994;
Charmandaris et al.\ 2000; Oosterloo \& Morganti 2005), this very close-by
system provides a unique laboratory for studying the jet--shell interaction
process. There, at two locations, the northern radio jet is seen being
interrupted by a gaseous shell and in each of those encounters the jet is
found to bend and flare-up towards the same (north-eastern) side, giving
rise to the two local radio peaks called the ``northern inner lobe" and the
``northern middle lobe" (GW10; GS84). As argued in these papers, such
repeated interruptions of the Cen A jets by the rotating gaseous shells at
different distances from the nucleus can explain the observed multi-peaked,
S-shaped, morphology of its radio lobes. Here we argue that such directly
observed jet-shell interactions might hold vital clues for understanding the
XRG phenomenon as well.  Shells have also been detected in the XRGs
3C 403 and 4C $+$00.58 (Almeida et al.\ 2011; Hodges-Kluck, et al. 2010b).
Unfortunately, since powerful radio galaxies are rare in the local
universe, all XRGs are sufficiently far away so that
the vast majority of shells that they may possess cannot be detected with current instrumentation.
However, given the frequency of their detection around nearby elliptical
galaxies that host RGs (e.g., Sikkema et al.\ 2007), it is most reasonable to investigate the hypothesis
that they are present in many such host galaxies and that 
suitably oriented jets will interact with them
as a matter of course.

Since the relativistic plasma jets in XRGs are typically $1 - 2$ orders of
magnitude more powerful than their counterparts in Cen A, they are unlikely
to be totally disrupted upon encountering the shells. It is much more
plausible that such encounters would slow down the jet and the lateral kick
imparted by the rotating shell would bend it sideways fairly abruptly in a fashion
somewhat similar to the situation seen in Cen A, as mentioned above. It is 
important to note that gas and dust have been detected in shells around
several other nearby elliptical galaxies (e.g., Schiminovich et al.\ 1995;
Balcells et al.\ 2001; Sikkema et al.\ 2007), while some cool gas has been
detected in a very large number of radio-AGN hosts (e.g., Sadler \& Gerhard
1985; Oosterloo et al.\ 2002) although it cannot be resolved sufficiently
to indicate that much of it is associated with shells. Thus it is quite reasonable to expect 
that more distant elliptical galaxies, hosting more powerful radio jets,
also possess gaseous shells which, however, cannot be detected with current 
instrumentation.

An additional impetus for our pursuing the jet--shell collision scenario for
XRGs comes from the  clear detection of a jet--cloud collision in the radio
galaxy 3C 321 at a redshift $z =$ 0.096 (Fig.\ 1 of Evans et al.\ 2008). It
is seen that the western jet of this powerful RG, while being impacted by the
ISM of a northward moving gas-rich galaxy, has undergone a sharp bend (by 
$\sim 40^{\circ}$). Both this deflected jet plume, which lacks 
a terminal hot spot, and the collimated main jet which has, post-interruption, 
resumed its original course and terminates in a radio/X-ray hot spot, are
presently visible in this source. Evans et al. have estimated that the jet's
interruption by the moving galaxy lasted for $<2 \times 10^6$yr.  Thus, within the XRG terminology,
the western half of this RG exhibits both an active (primary) radio lobe and
a relic secondary lobe (wing), closely resembling the XRG type radio structure.

Several powerful radio-loud quasars have long been known to 
show evidence for abrupt bends on large scales,
due to which they have been named ``dog-leg'' quasars and interactions
with dense clouds or even nearby galaxies have been suggested to explain
their morphology (Stocke, Burns \& Christiansen 1985).
We note that evidence for jet-cloud interactions on  smaller  scales is also available, 
e.g., the cases of the nearby Seyfert I galaxy III Zw 2 (Brunthaler et al.\ 2005)
and of the Broad Absorption Line quasar, QSO 1045$+$352 (Kunert-Bajraszewska et al.\ 2010). 

In this context, we stress that the key requirement of a large temporary 
deflection of a  light jet 
colliding with a few times wider massive cloud, along with subsequent stable 
flow, has been shown to be possible by  multiple groups performing 3D numerical 
simulations with different well tested hydrodynamics codes (de Gouveia Dal Pino 1999; 
Higgins, O'Brien \& Dunlop 1999; 
Wang, Wiita \& Hooda 2000).   Detailed simulations of light jets passing through many
small clouds show significant deviations to the jet flow, though not the large
bends that can be achieved by striking large clouds (Sutherland \& Bicknell 2007).
3D simulations of relativistic jets striking massive clouds also show structures similar
to that seen in 3C 321 forming under suitable circumstances (Choi, Wiita \& Ryu 2007)
and lasting for significant times. 
The remarkable observational  finding in 3C321, together with the two clear incidences of 
jet/shell collisions observed in Cen A (see above), taken together with the results of the 3D simulations, suggest the
following physically motivated scenario for the formation of  XRGs.

Usually, multiple shells are prominently observed within a bi-conical region
centered at the elliptical galaxy and having its axis broadly aligned with
the optical major axis of the elliptical (Malin et al.\ 1983), a result that
is consistent with computations of phase wrapping following a merger (Quinn 
1984). The existence of shells on opposite sides of the elliptical, forming
a bi-conical pattern broadly aligned with the elliptical's major axis, is 
particularly conducive for the success of XRG formation scenario based on 
jet-shell interaction. Such `aligned' systems of shells are preferentially 
found in elongated ellipticals (Prieur 1990 and references 
therein). Accordingly, only the radio jets emerging along the optical major 
axis would suffer significant interruptions by the shells. The resulting 
lateral impact on the jets, if adequate in strength and duration, would 
deflect them, giving rise to a pair of radio wings typically extruding at a 
large angle from the jet's original direction, as observed in 3C 321 (see 
above).

This observationally inspired scenario is capable of explaining the two main 
trends established for XRGs, namely, the strong tendency for their wings 
to be aligned with the optical minor axis of the host elliptical and for the
primary lobe axis to be roughly along the optical major axis (e.g., C02;
SS09; H-K10a). Simultaneously, it yields a simple explanation for the seemingly
puzzling result that the wings are often comparable, or even larger than 
the primary lobes (Sect.\ 2.2), which is an embarrassment for the backflow 
deflection model for XRGs (see below). As discussed in Sect. 2.2, Capetti
et al.\ (2002), in an attempt to address this formidable problem, have 
suggested that the driving mechanism for the backflow 
diversion is over-pressure of the radio cocoon, out of which a loosely 
collimated outflow of the synchrotron plasma would ensue along a direction
dictated by the buoyancy forces. Conceivably, the wings forming in this 
manner can grow (along the optical minor axis) as rapidly as the primary 
lobes (Sect.\ 2.2). However, as noted above, it is unclear if the radio 
lobes are in fact over-pressured (e.g., H-K10a). Moreover, as noted by 
these authors, in the well known XRG NGC 326, the ISM cannot be the 
`confining medium' required for the hypothesized over-pressured cocoon, 
since the northern secondary lobe originates well outside the ISM, although
if there is confinement by hot gas associated with a group or cluster then 
this gas might provide adequate confinement.  
Recall that in this XRG the secondary lobes again are a few times longer than 
the primary lobes (Worrall et al.\ 1995; Murgia et al.\ 2001).

 We wish to stress that, in this new model, if a post-merger RG has a jet 
orientation independent of the optical axis, as seems to be the case
for powerful sources (Battye \& Browne 2009), then only the random fraction 
of jets launched along the major axes are likely to encounter shells and 
thereby possibly form XRGs through this mechanism. The majority of jets would not end up being
intercepted by the shells and would hence evolve as normal RGs, as is of course 
observed, since XRGs are a small fraction of all RGs. 
If the preference for jets to roughly align with the optical 
minor axis, as seen in weaker RGs, is at all relevant for XRGs, whose powers 
tend to be near the borderline between the FR I and FR II sources, then 
this preference for the minor axis would imply that fraction of sources actually
aimed toward the major axes would be less than
expected from purely random jet/galaxy orientations.  
Still, we can speculate about a mechanism that would produce an observed correlation
between the jet direction and the galaxy's optical major axis, at least as 
seen in projection. If the incoming, smaller galaxy has no central SMBH, 
as appears to be the case with M33 and other dwarf galaxies (e.g., Ferrarese 
et al.\ 2006), then during the merger the spin of the SMBH in the primary galaxy would not 
be altered, nor would the  jet direction  reoriented. 
Nonetheless, if one averages over all possible directions, the random direction of 
a jet is expected to be within 30 degrees of any plane.   Of course as we 
can only see the entire configuration of the merged galaxies in projection, it therefore would not be 
surprising if the jet hits a shell.  

\begin{figure}[h!!!]
\centering
\includegraphics[width=9.0cm, angle=0]{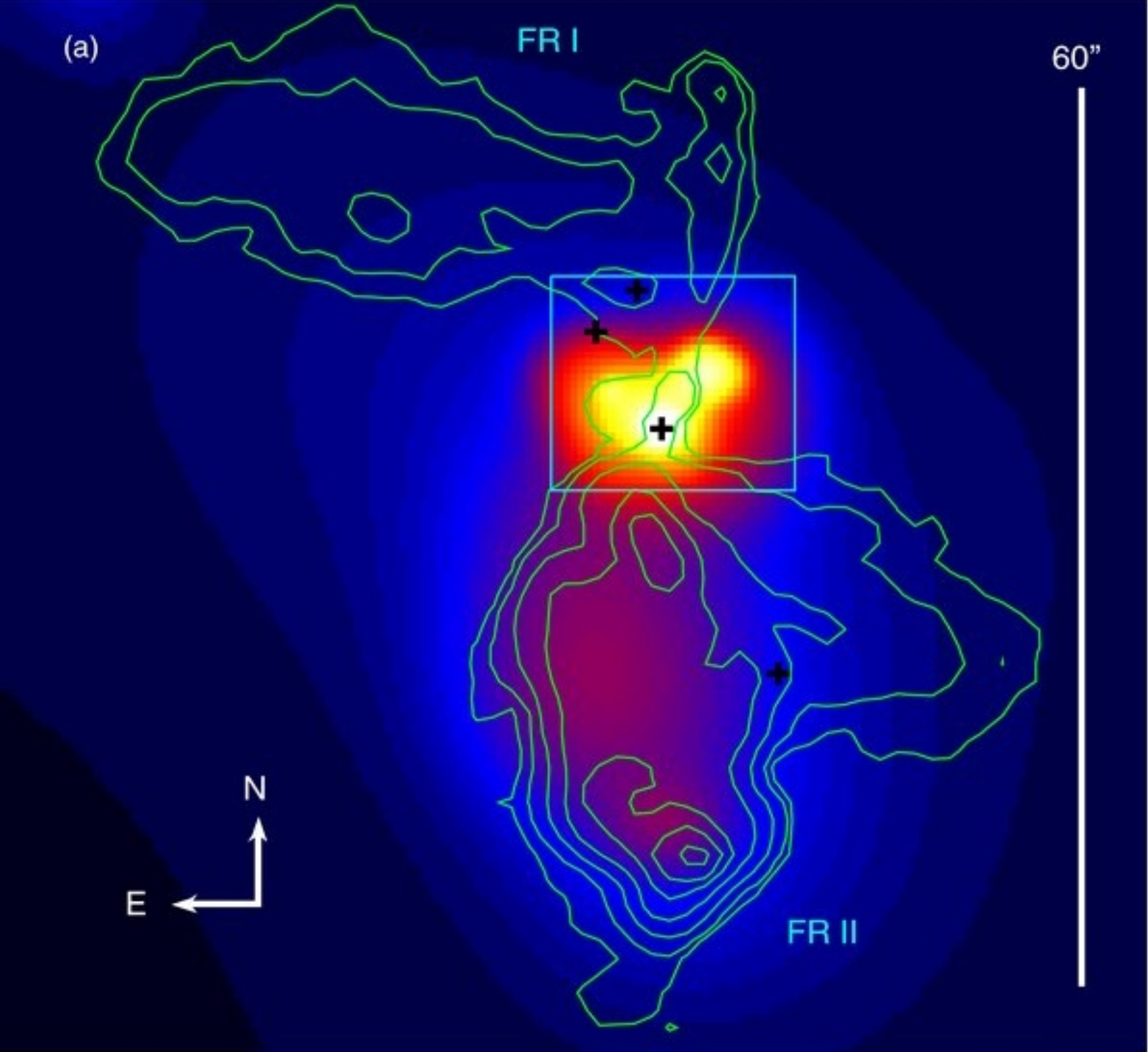}
\begin{minipage}[]{85mm}
\caption{Radio contours overlain on X-ray emission for the XRG 3C 433.  Reprinted 
by permission from Miller \& Brandt (2009);  copyright AAS}
\end{minipage}
\label{Fig1}
\end{figure}

As noted in Biermann \& Kronberg (1983) the extent of the outer fringes of
the ISM in early Hubble type galaxies can be quite large; however, the
pressure in the ISM strongly decreases outward and the pressure in
intra-group medium or even intra-cluster medium is very low. 
The observed near-orthogonal 
deflection of the two jets in opposite directions, giving rise to
the pair of secondary lobes, should occur naturally in the jet--shell
encounter model proposed here. Moreover, in this model the size of the
wings is determined not mainly by the speed of the backflow within the radio
lobes (vis-{\`{a}}-vis the speed of the hot spots). Instead, it would be
determined by the net duration over which the jet remained interrupted by the 
shell(s), as compared to the phase(s) during which the jet advanced 
uninterrupted by the shell(s).  In Fig.\ 1 a striking realization of the 
scenario proposed here can be seen in the recent VLA image of the RG 3C433
(Miller \& Brandt 2009: see also, Holt et al. 2007; Tadhunter et al. 2000).
>From the radio contours, the northern jet certainly appears to have resumed 
its advance recently, after successfully boring its way through a (hypothetical) 
shell which has caused a prolonged interruption of the jet that has quite plausibly 
resulted in the radio wing extending to the east.  Surviving evidence for the interruption
of the jet in the past is seen in the form of the compact radio peak (N3) precisely
at the point of the jet/shell interaction, which can be seen in this case
thanks to the sub-arc second resolution of the  VLA map by Black et al.\ (1992).
Such high quality maps are, unfortunately, not available for most XRGs so the signatures of
jet/shell interactions are usually less clear as it is the case for nearby radio
galaxies such as 3C 433 and Cen A.

Thus, this simple picture can readily allow for the possibility of secondary 
lobes (wings) being more extended than even the primary lobes (GW10), as 
observed in several XRGs (e.g., 3C 223.1 and 3C 403, Dennett-Thorpe et al.\ 
2002; NGC 326, Murgia et al.\ 2001; J1130+0058, Zhang et al. 2007; 4C+00.58, 
Hodges-Kluck et al. 2010b). At the same time, it is clear that even if jets'
interuptions by the shells (disposed roughly along the optical major axis) are 
not strong enough to cause significant diversion of their flow, the repeated
stalling of the jets would retard their advance. This could well be why giant
radio sources are preferentially seen to align with the optical minor axis of 
the host galaxy, as found by SS09, Palimaka et al.\ (1979) and Guthrie (1980). 
Lastly, the observed Z-shaped morphology of the secondary lobe pair (GBW03) 
about the nucleus is clearly another natural outcome of this scenario. We note 
that this assertion would hold even if the Z-deflections of the two jets occur 
at vastly different separations from the nucleus, as for instance seen in the 
XRG NGC 326 (see above). This is because, in this scenario, the deflection 
points of the two jets are determined by the radial locations of the two 
(partial) shells, on the opposite sides of the galaxy along its major axis, 
which were successful in effectively interrupting and deflecting the respective 
jets.

The jet-shell interaction model proposed here for XRGs may also be invoked
for understanding  another intriguing recent results about powerful radio 
galaxies. It has been noted that the lobes of RGs display more pronounced 
lateral distortions when the radio source axis is better aligned with the 
major axis of the host elliptical (SS09). This can be readily understood 
since the interaction of the lobe plasma with the shells that causes the lobes'
distortion would happen mostly when the lobes happen to lie along the host's 
major axis, i.e., where the shells preferrentially form.
Finally, the jet-shell collision scenario for the formation of wings may 
even apply in the exceptionally rare case of the XRG 4C$+$00.58, where the 
jet appears aligned with the optical minor axis, provided a gradual jet 
reorientation via accretion, as inferred by Hodges-Kluck et al.\ (2010b), 
is a valid description of the system. Bearing in mind the vast potential 
of this scenario for explaining several intriguing morphological aspects 
of double radio sources, we strongly encourage deep optical searches for 
the shells about the host galaxies of XRGs, whose counterparts in several 
nearby ellipticals were discovered already three decades ago (Malin \& 
Carter 1980, 1983; Malin et al.\ 1983) and whose relevance to the intriguing 
RG morphologies was recognized soon thereafter (Gopal-Krishna \& Chitre 1983; 
GS84). In this context, it is encouraging that a rich system of shells has
recently been discovered in the archetypal XRG 3C403 (Almeida et al.\ 2011).
Likewise, it is also important to establish the nature of the environment 
around XRGs; in the jet-shell interaction picture they are expected to reside 
in comparatively low density environment, like the shell galaxies. 
It may also be noted that even on parsec scale jets seem to get repeatedly 
interrupted by gas clouds, as revealed by the extensive VLBI observations of 
the nearby Seyfert I galaxy III Zw 2 (Brunthaler et al.\ 2005).

\section{Summary and Main Conclusions}

In the foregoing sections we have endeavoured to collate various available
clues in order to arrive at an improved understanding of the mechanism 
responsible for the X-shaped radio galaxies (XRGs). Thus, we confronted the
various proposed models with a fairly large body of observations available now. 
Although rather enigmatic, XRGs may not be rare objects, when it is realized 
that the XRG classification demands not merely that the primary and secondary 
lobe axes have a large misalignment, but also that they are oriented far from 
the line of sight. A condensed overview of the principal merits and shortcomings 
of the various proposed models is presented in Table 1. 

While an occasional XRG might arise from the superposition of two
independent AGN inside a merging pair of ellipticals this explanation 
does not accord with the bulk of the data. A simple buoyancy aided
back-flow diversion model (e.g., LW84) might be adequate to explain 
RGs with modest ``wings'', where the secondary lobes are short compared 
the primary.  The only way this class of models could account for the 
not-infrequent occurence of secondary lobes comparable or even larger 
than the primary lobes, is for the backflow filled cocoon to become 
strongly over-pressured, so that the relativistic plasma might rapidly
squirt out and escape along the minor axis of the host galaxy. If then, the 
primary jets happen to be fortuitously oriented along the host elliptical's 
major axis, the interesting correlation found by Capetti et al.\ (2002) 
can probably be understood. However, the numerical simulations purporting 
to demonstrate this is feasible  were long restricted to two-dimensions (C02), 
and this artificially induced symmetry is very likely to exaggerate the 
collimation and strength of these outflows.  However, very recent 3-D simulations
have demonstrated that it is possible to obtain  lengthy overpressured outflows
if the host galaxy has a very high ellipticity (Hodges-Kluck et al.\ 2011); 
if viewed at particularly favorable orientations the secondary lobes can even
appear to be longer than the primary ones in projection (Hodges-Kluck et al.\ 2011;
Wiita et al.\ 2011).

In addition, the over-pressured cocoon model does not seem capable of 
explaining the important Z-symmetric subset of XRGs. Moreover, convincing 
evidence for the requisite over-pressured regions within the cocoons is 
still lacking. There is a strong evidence that mergers are necessary for
triggering at least some AGN and several nearby AGN, such as Cen A, are 
prime examples of an elliptical having rather recently swallowed a gas and 
dust rich neighbour. These mergers are of great interest also for the 
generation of gravitational radiation and perhaps prompt electromagnetic 
signals as well. 

The basic spin-flip model also can readily explain why at most only one pair of 
lobes is ever of the FR II type. In addition, a simple
variant of this basic picture is able to produce the Z-shaped distortion
clearly witnessed in some of the XRGs, another important merit. Thus it seems
likely that at least some XRGs are indeed created in this manner. The spin-flip 
model also produces very interesting and specific relations between 
gravitational radiation and electromagnetic wave signatures (e.g., GB09), 
and their detection would be extremely important and also would provide proof of 
the relevance of such mergers in AGN.
However, this physically well motivated spin-flip model offers 
no natural explanation for the correlation of the primary lobe pair with the
optical major axis of the parent elliptical,
and so probably  cannot account, by itself, for the majority of XRGs.   

Related to the spin-flip scenario we show in the Appendix A that the probability of any specific mass ratio $q$ is close to a flat distribution in $\Delta q/q$; thus mass ratios close to unity should be exceedingly rare. We also address a number of observations which occur for semi-major mergers (defined as having the most likely mass ratios, between 1:30 to 1:3, 1/10 being typical). In Appendix B, we compute the spin evolution of the radio-loud (primary) SMBH as well as the expected intervals between successive semi-major BH mergers. We find that the spin reorientations during the two phases (separated by the time when the declining orbital momentum of the infalling SMBH has become comparable to the spin momentum of the main SMBH) are similar in magnitude, but the rate of spin evolution is about $10^4$ times faster in the second phase, resulting in a duration as short as $\sim$ 3 years, while the precession time-scale is less then a day. If the rapidly rotating jets are relativistic and if they came close to our line of sight, they could produce significant variability at all wavelengths (but mostly detectable in the radio, hard X-ray and gamma-ray) years before the coalescence. Such systems would provide a detectable EM precursor of the {\it impending} strong GW emission.

A careful scrutiny of the mechanisms leading to XRGs shows that while each of them might
play a dominant role in different XRGs, the mechanisms involving galaxy
mergers, which include the spin-flip scenario, are likely to have the widest 
applicability, considering their greater overall compatibility with the 
observations. More specifically, the various observational constraints in 
some sources have led us to propose a jet-shell interaction model for XRGs. 
While this proposed new alternative mechanism also involves a galaxy merger 
and assumes that each galaxy possesses a SMBH, as does the spin-flip model, 
it does not require the galaxy to have been radio loud prior to the merger.
However, it may be noted that a spin-flip will typically have occurred 
in the course of merger and the direction of the jets thus triggered 
will probably be along the axis defined by the angular momentum of the 
accreted galaxy and therefore also perpendicular to any central dust-lane. 
The powerful interruptions of jets by moving massive clouds in the environments
of the host galaxy, as observed in some nearby RGs, such as Cen A and 3C 321, 
provide fairly persuasive evidence in favour of the jet-shell interaction 
scenario being important in a goodly subset of XRGs.
Here it is relevant to recall that the stellar shells (which are probably 
also gaseous) in a bi-conical pattern are preferentially disposed along the 
optical major axes of the 
post-merger galaxy. When these features are combined with simulations that 
show substantial sideways flows from powerful jets striking massive clouds, 
the wide range of morphological properties associated with XRGs 
can be understood.

\begin{figure}[h!!!]
\centering
\includegraphics[width=14.0cm, angle=0]{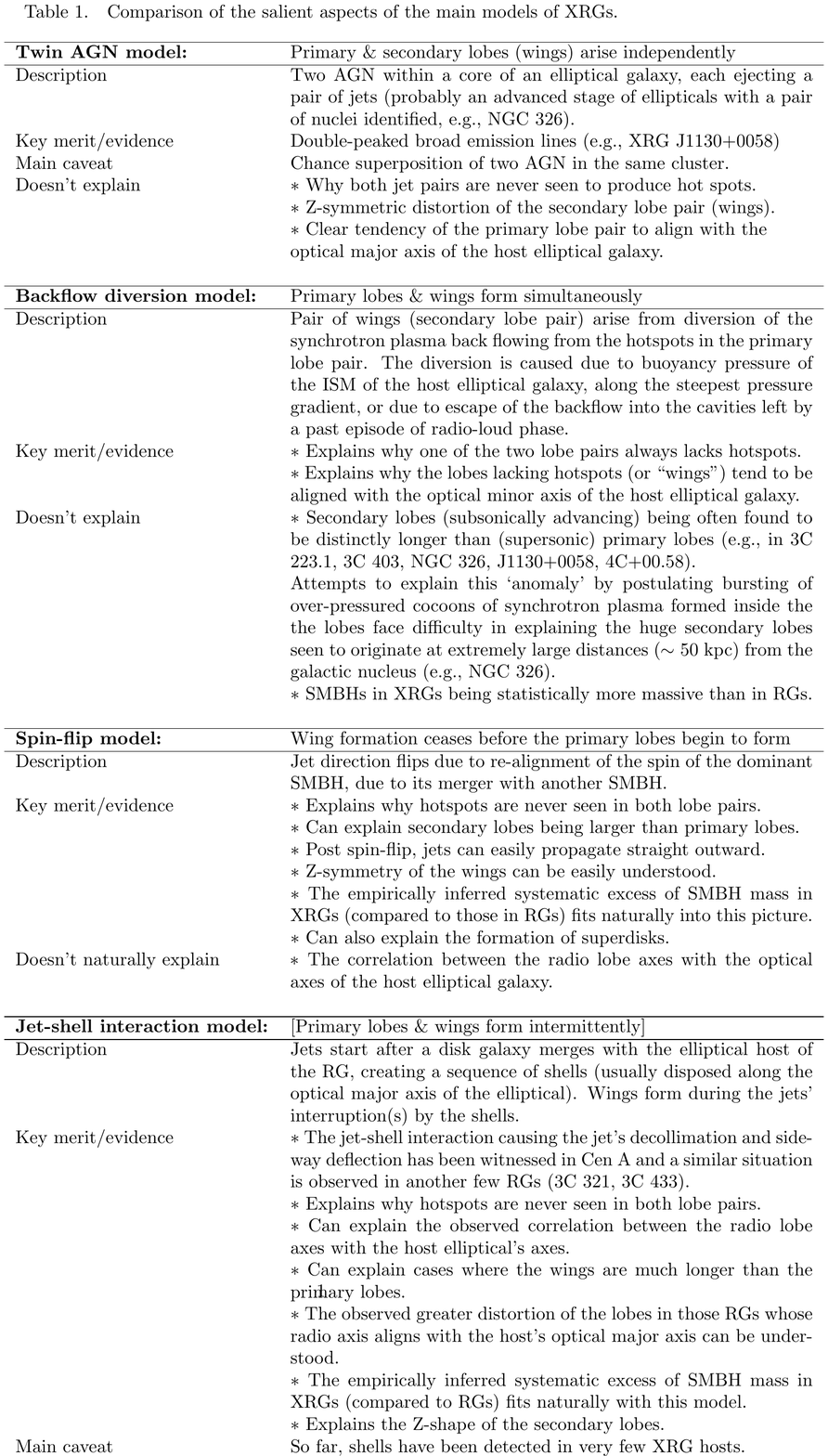}
\label{Fig2}
\end{figure}

\section{Acknowledgments}

We thank the referee for pointing out important additional references and for suggestions
that clarified the presentation.
PLB wishes to thank Lakshmi Saripalli, Chanda Jog and Elisabete de Gouveia Dal Pino for extended  discussions and gratefully acknowledges the award of a Sarojini Damodaran Fellowship (2009).
Support for work by PLB has come from the AUGER membership and theory grant 05 CU 5PD 1/2 via DESY/BMBF and VIHKOS. 
L\'{A}G is grateful to Tiberiu Harko for kind hospitality during his visit at the University of Hong Kong. He was partially supported by the
Hungarian Scientific Research Fund (OTKA) grant no.\ 69036 and by COST Action MP0905 "Black Holes in a Violent Universe". 
The work of PJW was supported in part by a subcontract to US
NSF grant AST 05-07529 to the University of Washington.

\appendix

\section{The expected mass ratio of merging SMBHs and consequences}

In GB09 the characteristic merger ratio was estimated to be between 3:1 to 30:1, the value of 10:1 being found typical. Here we give independent considerations supporting the statement that both equal mass encounters and extreme mass ratio mergers, which qualify as a test particle falling into a black hole, are rare. 

Mergers between two black holes both drawn from a black hole mass distribution
can be written as integrals over a merger rate $\alpha(M_{1}, M_{2})$
multiplied by the density at each of the masses $M_{1}$ and $M_{2}$.  This
approach uses the assumption that the highest density region of the
universe are in groups and clusters, where the density depends (actually does not depend)
on redshift as $(1+z)^{0}$ (Cavaliere et al. 1991, 1992, 1997) up to
some critical redshift.  
Silk \& Takahashi (1979) have found an analytic asymptotic solution to the merger equation using a merger rate
running as $M_{1}^{\epsilon/2} M_{2}^{\epsilon/2}$.  We can reproduce the data to a good approximation with a
merger rate varying as $M_{1}^{2/3} \, M_{2}^{2/3}$; in fact, this exponent, $\epsilon = 4/3$, is the gravitational
focusing limit in the analysis of Silk \& Takahashi.  This gives a low mass tail of $M^{-2}$ for the BH mass
distribution, and a high mass tail, which can be approximated by $M^{-3}$, quite close what the data show (Caramete
\& Biermann 2010).
These two regimes are joined at a break mass $M_{b}$ of about
$10^{8} \; M_{\odot}$.  

Including mergers between two black holes both
below the bend, 60 percent of all mergers occur within the mass ratio ranges 3:1 and 100:1. Ignoring mergers between two partner black holes both below
the break mass we work out the frequency in various mass ratio ranges in
more detail: for the ratio ranges $q=$ 1 to 3, 3 to 10, 10 to 30, and 30
to 100, the corresponding percentages are 3, 15, 61 and 21 percent.
Therefore the most common mass ratio is 10 to 30 with 61 percent of all mergers,
and an additional 36 percent of mergers occur in the two adjoining mass ranges. Therefore, the mass ratio range from 3:1 and 100:1 covers 97 percent of all mergers. 

Under specific initial conditions the linear momentum kick imparted to the merged SMBHs can exceed several thousand km s$^{-1}$, which is enough for the merged SMBH to escape from the host elliptical (Bogdanovi\'{c}, Reynolds \& Miller 2007). These large linear momenta can only be produced when the BH mass ratio is nearly 1, and for restricted spin configurations (anti-aligned spins, lying in the orbital plane). We will argue below that such initial conditions are unlikely. First, as seen above, the probability of any specific mass ratio $q$ is close to a flat distribution in $\Delta q/q$; thus mass ratios close to unity should be exceedingly rare.

Secondly, any gas/circumbinary disk will act toward an alignment of the spins with the orbital momentum, significantly reducing the kick velocity (Bogdanovic et al. 2007). This occurs through the Bardeen-Petterson effect (Bardeen \& Petterson 1975), which in turn is based on the Lense-Thirring precession. The real situation is expected to fall between the two rare extremes, namely, an extremely `wet' merger (i.e., copious amount of gas associated with it, leading to this pre-merger realignment), and a completely `dry' merger with essentially no gas (hence allowing for random spin orientations). 
Therefore it is expected that even the presence of some gas or a circumbinary disk will disfavour the spin configuration leading to a large recoil, but still leave room for a significant spin-flip. 

Consequent to a `semi-major' merger the SMBH spin could undergo a large 
directional `flip'. The new spin axis would tend to get reoriented towards the orbital momentum axis of the in-spiraling massive BH (and, therefore, the ISM rotation axis: see GB09). Such a geometry would ensure, firstly, that the post-merger jets, ejected along the new spin axis, will encounter little lateral impact from the rotating ISM and hence propagate essentially straight outward. Secondly, the old lobes would gradually acquire the appearance of wings/relics which would be kept energized at a significant level as the synchrotron plasma backflowing from the new hot spots would find a natural escape sideways into these pre-existing low-pressure relic lobes (GBW03; LW84). 

Following a semi-major merger (SMBH mass ratio $>1:30$) the spin axis of the coalesced SMBH pair is expected to settle down roughly
along the direction of the original orbital angular momentum of the
secondary galaxy about the primary one (GB09). Then the same axis is likely to be retained during any subsequent ejections of restarted jets. This is because the probability of recurring semi-major mergers, each causing a large spin-flip is quite small, in that the temporal interval between them certainly is $>1$ Gyr for semi-major mergers with mass ratios $>1:30$ and even longer for major mergers (Z.\ Lippai, private communication, based on the merger tree program in Lippai et al.\ 2009). Whereas, if jets are restarted while the earlier lobes are still visible, then it is likely that they have done so within $\sim 10^{8}$ yr of the launch of the original jets, since visible radio activity seldom lasts much longer than that (e.g., Blundell \& Rawlings 1999; Gopal-Krishna \& Wiita 2001; Barai \& Wiita 2007; Kaiser \& Best 2007). In this way, the situations where the incipient inner lobe pair is essentially aligned with the (outer) primary lobes, as shown for a few XRGs (SS09), can be explained naturally. An alternative possibility for such an observed alignment is from a fortuitous projection of a real XRG onto the plane of the sky in such a way that the shorter primary lobes appear embedded within the decaying secondary lobes, thereby giving the appearance of a ``double-double'' source (Wiita et al.\ 2011).

\section{The expected time evolution of the SMBH spin}

In GB09 the following evolution of the inclination angle was derived%
\begin{equation}
\dot{\alpha}\approx \frac{32c^{3}}{5Gm}\varepsilon ^{9/2}\eta \nu
^{-1}\left( \frac{L}{J}\right) ^{2}\sin \left( \alpha +\beta \right) ~,
\label{alphadotepeta}
\end{equation}%
with $\varepsilon =Gm/c^{2}r$ the post-Newtonian parameter (increasing
during the inspiral with the decreasing separation $r$; here $m=m_{1}+m_{2}$
is the total mass), $\alpha $ and $\beta $ the angles spaned by the orbital
angular momentum and the spin with the total angular momentum (their sum, $%
\alpha +\beta $ being constant during the inspiral), $\nu =m_{2}/m_{1}\leq 1$
is the mass ratio, $\eta =m_{1}m_{2}/m^{2}=\nu \left( 1+\nu \right) ^{-2}=\mu
/m\,$,  the symmetric mass ratio and $J$ the magnitude of the total angular
momentum. For mutually perpendicular spin and orbital angular momentum  $%
J^{2}=L^{2}+S_{1}^{2}$.   For circular orbits and to
leading order, $S_{1}/L=\varepsilon ^{1/2}\nu ^{-1}\chi _{1}$ holds (where $%
\chi _{1}\in \left[ 0,1\right] $ is the dimensionless spin), and we now obtain 
\begin{equation}
\left( \frac{L}{J}\right) ^{2}=\left( 1+\varepsilon \nu ^{-2}\chi
_{1}^{2}\right) ^{-1}~,
\end{equation}%
thus the evolution of the inclination angle is described by 
\begin{equation}
\dot{\alpha}\approx C\nu ^{-1}\varepsilon ^{9/2}\left( 1+\varepsilon \nu
^{-2}\chi _{1}^{2}\right) ^{-1}~,  \label{alphadot}
\end{equation}%
with the constant coefficient given as%
\begin{equation}
C=\frac{32c^{3}}{5Gm}\eta \sin \left( \alpha +\beta \right) ~.
\end{equation}%
We note that $\dot{\alpha}$ as a function of $\varepsilon $ is monotonically
increasing, thus the spin-flip rate increases with increasing post-Newtonian
parameter (with decreasing radius).

In order to find $\delta \alpha $ during any part of the inspiral we
have to integrate 
\begin{equation}
\delta \alpha =\int \dot{\alpha}\left( r\right) dt=\int \frac{\dot{\alpha}%
\left( r\right) }{\dot{r}_{gw}}dr~,
\end{equation}%
where by $\dot{r}_{gw}$ we mean the rate of the inspiral. As for a Keplerian
circular orbit $L^{2}=Gm\mu ^{2}r$, therefore $\dot{r}_{gw}=2\left( Gm\mu
^{2}\right) ^{-1}L\dot{L}^{gw}$. Employing 
\begin{equation}
\dot{L}^{gw}=-\frac{32G\mu ^{2}}{5r}\left( \frac{Gm}{c^{2}r}\right) ^{5/2}~,
\end{equation}%
which holds on the radial orbit in average (Apostolatos et al.\ 1994), we get%
\begin{equation}
\dot{r}_{gw}=-\frac{64\eta }{5}c\varepsilon ^{3}~,
\end{equation}%
thus%
\begin{equation}
\delta \alpha \approx \frac{5C}{64\eta \nu c}\int \frac{\varepsilon ^{-1/2}}{%
1+\varepsilon \nu ^{-2}\chi _{1}^{2}}d\varepsilon ~.
\end{equation}%
For mass ratios about $\nu \approx 0.1$ the inspiral begins at about $%
\varepsilon ^{\ast }\approx 10^{-3}$ (GB09), it proceeds through the $L\gg
S_{1}$ phase until $L\approx S_{1}$ is reached (when $\varepsilon \approx
\nu ^{2}$), then continues through the phase $L\ll S_{1}$ until $\varepsilon
_{fin}=\left( Gm/c^{2}r_{ms}\right) $, where $r_{ms}$ is the marginally
stable orbit. For high spin ($\chi _{1} \sim 1$) we have $r_{ms}\sim
2Gm/c^{2}$ thus $\varepsilon _{fin}\sim 0.5$. Let us estimate the ratio
of the changes in the inclination angle (the ratio of spin-flips) during
these two phases. We obtain%
\begin{equation}
\frac{\delta \alpha _{I}}{\delta \alpha _{II}}=\frac{\int_{10^{-3}}^{10^{-2}}%
\frac{\varepsilon ^{-1/2}}{1+100\varepsilon }d\varepsilon }{%
\int_{10^{-2}}^{0.5}\frac{\varepsilon ^{-1/2}}{1+100\varepsilon }%
d\varepsilon }=0.74~.
\end{equation}%
Therefore, the respective spin-flips are comparable. 

Now let us see how long the phases last. For this we evaluate in a similar
way%
\begin{equation}
\delta t=\int_{t_{1}}^{t_{2}}dt=\int_{r_{1}}^{r_{2}}\left\vert \frac{dr}{%
\dot{r}_{gw}}\right\vert =-\frac{5Gm}{64\eta c^{3}}\int_{\varepsilon
_{1}}^{\varepsilon _{2}}\varepsilon ^{-5}d\varepsilon =\frac{5Gm}{256\eta
c^{3}}\left. \varepsilon ^{-4}\right\vert _{\varepsilon _{1}}^{\varepsilon
_{2}}~,
\end{equation}%
to see that the time-scale varies linearly with the total mass.  So then we have  
\begin{equation}
\frac{\delta t_{I}}{\delta t_{II}}=\frac{\int_{10^{-3}}^{10^{-2}}\varepsilon
^{-5}d\varepsilon }{\int_{10^{-2}}^{0.5}\varepsilon ^{-5}d\varepsilon }%
\approx 9999~.
\end{equation}

We conclude that the second part of the spin-flip occurs approximately $%
10^{4}$ times faster than the first part. Due to the increased
post-Newtonian parameter and increased rate of change of the inclination
angle, a considerable and certainly observable part of the spin-flip will
occur during the last stages of the inspiral. An estimate for this was given
in GB09 as 3 years.

\end{document}